\documentclass[twocolumn]{revtex4-2}
\usepackage{amsmath}
\usepackage{graphicx}
\usepackage{booktabs}
\usepackage{subcaption}
\usepackage{xspace}
\usepackage{multirow}
\usepackage{float}
\usepackage{siunitx}
\usepackage{geometry}
\usepackage{empheq}
\begin{document}

\title{Multi-Dimensional Phase Space Manipulation for Attosecond Electron Bunch Compression}

\author{Yuxin Cheng}
\affiliation{Shanghai Institute of Applied Physics, Chinese Academy of Sciences, Shanghai 201800, China
University of Chinese Academy of Sciences, Beijing 100049,China
}

\author{Qiang Gu}
\affiliation{Shanghai Advanced Research Institute, Chinese Academy of Sciences, Shanghai 201210, China}

\author{Chao Feng}
\email{fengc@sari.ac.cn}
\affiliation{Shanghai Advanced Research Institute, Chinese Academy of Sciences, Shanghai 201210, China}

\begin{abstract}
Attosecond electron beams are essential for investigating ultrafast structural and electronic dynamics in matter with atomic-scale resolution. We propose a novel method that enables robust attosecond-level electron bunch compression. This method employs THz-driven linear energy chirping and multidimensional phase-space manipulation, effectively compressing the electron bunch and suppressing its arrival timing jitter. Implemented in an MeV ultrafast electron diffraction beamline, this method compresses a 3~MeV, 0.1~pC electron beam from an initial duration of 50~fs to 810~as while retaining 6~fC of charge, with 850~as arrival-time jitter. This approach enables unprecedented timing resolution in ultrafast sciences and offers significant potential for other accelerator applications involving attosecond-scale electron beams.
\end{abstract}
\maketitle

\section{Introduction} 
Electron beam-based ultrafast techniques, particularly ultrafast electron diffraction (UED), has become a vital tool for investigating atomic-scale structural dynamics with femtosecond temporal and ångström spatial resolution~\cite{PotentiFemtosecond,Highued,weathersby_mega-electron-volt_2015,Hydrogen_positions,10.1063/1.4926994,nanolett.5b02805,yang_diffractive_2016}. The temporal resolution of UED is fundamentally constrained by the duration and shot-to-shot arrival-time jitter of the electron bunch~\cite{srinivasan_ultrafast_2003}. Early implementations of UED relied on photocathode-driven electron guns operating at tens to hundreds of kiloelectronvolts~\cite{nie_femtosecond_2009,srinivasan_ultrafast_2003,sciaini_electronic_2009}. These nonrelativistic beams experienced severe pulse broadening due to space charge force and temporal dispersion during DC acceleration, imposing a hard limit on achievable time resolution~\cite{williamson_clocking_1997,nie_femtosecond_2009}.

In recent years, extensive efforts have been devoted to pushing the temporal boundary of UED. Photocathode radiofrequency (RF) guns at MeV energies took advantage of relativistic mitigation of space-charge forces, enabling generation of much shorter pulses~\cite{musumeci_high_2010,Highued}. Conventional RF compression approaches that imprint an energy chirp via cavity phase modulation and use velocity bunching in drift sections can yield sub-20 fs pulses but introduce tens of femtoseconds of arrival-time jitter due to RF phase noise~\cite{PhysRevLett.118.154802,PhysRevX.8.021061}. Laser-driven THz bunchers eliminate phase jitter through intrinsic laser-THz synchronization, yet residual RF amplitude instabilities and downstream beamline dispersion convert energy fluctuations into 70 fs of timing jitter~\cite{PhysRevLett.124.054802,PhysRevLett.124.054801}.

Double-bend achromat compressors address RF amplitude sensitivity by engineering isochronous transport, remarkably reducing timing jitter to a few femtoseconds~\cite{qi_breaking_2020}. Nevertheless, state-of-the-art compression methods still cannot reach the attosecond regime, because higher-order nonlinearities in the longitudinal phase space induce pulse broadening, and aggressive collimation to suppress space charge tails drastically reduces usable beam charge. These challenges highlight that two-dimensional electron beam manipulation in the longitudinal phase space can hardly simultaneously achieve sub-femtosecond durations, suppress timing jitter, and preserve high bunch charge, underscoring the urgent need for a new compression paradigm.

Here, we introduce a multi-dimensional phase-space manipulation approach that achieves attosecond-level compression of MeV electron beams while preserving beam quality. By combining pulse-front-tilt laser-induced transverse-to-longitudinal coupling at the photocathode, THz-driven linear energy chirping immune to RF jitter, and a specially designed beamline that transforms transverse emittance into longitudinal compression, this method simultaneously compresses the electron bunch to attosecond durations, decouples arrival time fluctuations from RF amplitude noise, and preserves bunch charge.

\section{Methods}
\label{sec:model}

\begin{figure*}[htbp]
    \centering
    \includegraphics[width=\textwidth]{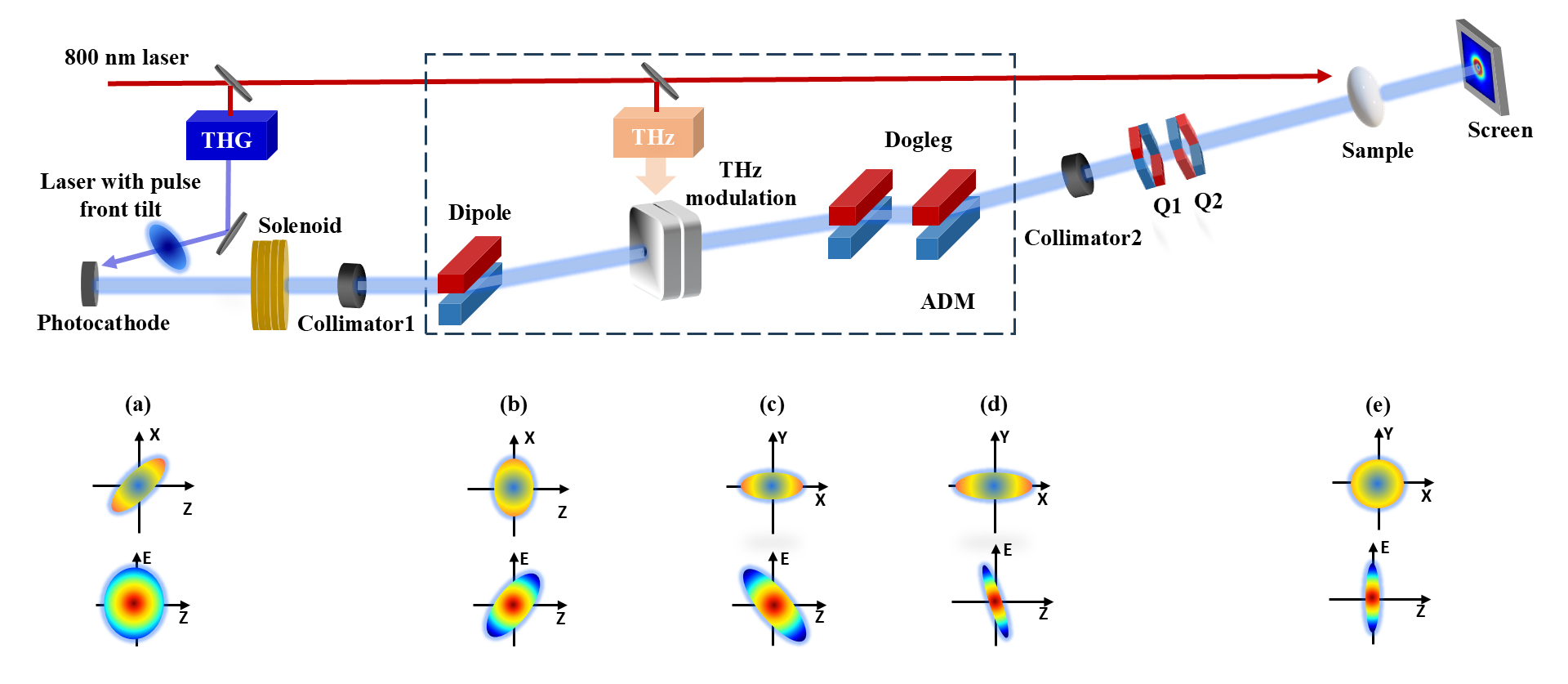}
\caption{Schematic of the multi-dimensional phase-space manipulation beamline. A pulse-front-tilt UV laser illuminates the photocathode to generate the electron beam. The beam is then accelerated in the RF gun, with its transverse size controlled by a solenoid lens, before entering the ADM beamline, where transverse divergence is converted into longitudinal density compression. Skewed quadrupole doublets (Q1, Q2) subsequently restore symmetric transverse emittances. The multi-dimensional phase space evolution of the electron beam from the cathode to the sample are illustrated in (a-e).}
\label{fig:adm_workflow}
\end{figure*}

The schematic layout of the proposed method is illustrated in Fig. 1. Here we adopt the Angular Dispersion-Induced Microbunching (ADM) scheme~\cite{feng_storage_2017} to perform the required transverse-to-longitudinal emittance partitioning. The ADM configuration comprises three stages: first, a magnetic dipole imparts angular dispersion to the electron beam; second, a modulator introduces a controlled energy chirp; and third, a dogleg compressor applies both transverse and longitudinal dispersion to compress the beam. The ADM was initially developed to generate periodic microbunchings in a long electron bunch through precise control of transverse-longitudinal phase space correlations. The final duration for an individual microbunch is set by the initial angular divergence of the electron beam, much shorter than what is attainable with energy modulation alone.

When the electron beam passes through the first dipole of ADM, the intrinsic transverse-longitudinal coupling stretches the bunch by several tens of femtoseconds. While this elongation is negligible for longer pulses, it induces nonlinear distortions during subsequent THz-driven energy modulation of ultrashort femtosecond bunches, ultimately degrading the compression quality. To mitigate this effect, we employ a pulse-front–tilted laser to illuminate the photocathode (Fig. 1), generating the required correlation at emission. Such pulse-front tilt can be realized either by oblique laser incidence or by guiding the beam through a double-grating assembly that imparts transverse dispersion.

The sequence begins with the tilted-pulse illumination shown in the upper panels of Fig. 1a. The pulse-front tilt creates a one-to-one mapping between the electrons’ longitudinal positions and their transverse coordinates. After passing through the photocathod RF gun and a dirft section with longitudinal dispersions of \(R_{56}^{\rm gun}\) and \(R_{56}^{\rm drift1}\), respectively, the beam was sent into the ADM section. After passing through the first dipole with bending angle of \(b\), the x-z-coupling induced by the drive laser has been counteracted (Fig. 1b, upper panels), thereby suppressing nonlinear distortions in the following THz modulation. In the longitudinal phase-space view (lower panels of Fig. 1a), the beam initially acquires a positive energy chirp, meaning the head carries higher energy than the tail, due to space-charge forces in the gun. As the bunch drifts with negative longitudinal dispersion, the faster head moves ahead, lengthening the pulse (Fig. 1b, lower panels). The THz modulator then over-compensates this chirp, reversing it to a negative value \(h = \Delta\gamma/\gamma_0 \cdot k_s\), where \(k_s\) is the frequency of the THz-induced modulation (Fig. 1c, lower panels). The subsequent dogleg with Longitudinal dispersion \(\xi_{D}\) and transverse dispersion \(\eta\) converts this energy modulation and transverse emittance into longitudinal compression (Fig. 1d and 1e, lower panels). After that, the electron beam passages through another drift section with longitudinal dispersions of \(R_{56}^{\rm drift2}\) before the sample. By tuning the dispersive properties of the first dipole, dogleg, and final drift section, the contribution of the initial energy spread to the final bunch duration can completely vanished and an ultrashort bunch can be achieved. In the transverse plane, the combined dispersions of the dipole and dogleg produce an asymmetric beam profile after the ADM section due to emittance anisotropy (Fig. 1d, upper panels). To restore a round beam, we introduce a skewed quadrupole doublet downstream of the ADM line. This element redistributes the transverse phase space and transforms the flattened profile into a symmetric, round beam (Fig. 1e, upper panels)~\cite{dowell_exact_2018}.

The total longitudinal dispersion \(R_{56}^{\text{total}}\) from the cathode to the sample should be engineered to satisfy the isochronicity condition:
\begin{equation}
\begin{aligned}
    R_{56}^{\text{total}}=R_{56}^{\text{gun}} &+ R_{56}^{\text{drift1}}+R_{56}^{\text{ADM}} + R_{56}^{\text{drift2}} = 0,
\end{aligned}
\label{eq:isochronicity}
\end{equation} 
where \(R_{56}^{\rm ADM}\) represents the longitudinal dispersion of the ADM section. The isochronicity condition ensures that arrival time jitter remains decoupled from RF amplitude fluctuations. And the final longitudinal position of the electron is

\begin{equation}
z_{\text{final}} = 
\begin{aligned}[t]
&\underbrace{b(1 + h\xi_D)x}_{\substack{\text{Bending-induced} \\ \text{displacement}}} + \underbrace{\eta x'}_{\substack{\text{Compression} \\ \text{term}}} \\
&+ \underbrace{(1 + h\xi_D)z}_{\substack{\text{Chirped} \\ \text{drift}}} + \underbrace{(R_{56}^{\text{ADM}} + R_{56}^{\text{drift2}} - \eta b)\delta,}_{\substack{\text{Isochronicity} \\ \text{correction}}}
\end{aligned}
\label{eq:z_final}
\end{equation}
where \(x\) and \(x'\) is the horizontal position and divergence, respectively, \(z\) is the initial longitudinal position, and \(\delta=\Delta \gamma/\gamma_0\) is the relative energy deviation. Under the optimized condition of ADM, all the terms in the parentheses of the above equation can be zero, and we get

\begin{equation}
\begin{aligned}
z_{\text{final}} = \eta x',\
\end{aligned}
\label{eq:z_final1}
\end{equation}
indicating that the final duration and timing jitter of the electron beam are mainly determined by the angular divergence, which can be very small and stable after the collimator1 upstream of the ADM section.

\section{Simulations}
\label{sec:model}

To demonstrate the performance of the proposed method, three-denominational simulations with GPT (General Particle Tracer)~\cite{GPT} has been perfromed based on the design criteria defined in Eqs. (1) and (3). The simulation parameters are summarized in Table~~\ref{tab:full_params}. Initial electron beam conditions included a kinetic energy of 3~MeV, 0.1~pC charge, and 50~fs pulse duration.

\begin{table}[htbp]
\centering
\caption{ Simulation Parameters}
\label{tab:full_params}
\begin{tabular}{lcc}
\toprule
\textbf{Parameter} & \textbf{Value} & \textbf{Unit} \\
\midrule
\multicolumn{3}{l}{\textit{Beam Initial Conditions}} \\
\hline
Kinetic energy & 3.18 & MeV \\
Charge & 0.1 & pC \\
Initial bunch length& 50 & fs \\
normalized emittance & $1\times10^{-2}$ & mm·mrad \\
Energy spread & 0.1 & \% \\

\midrule
\multicolumn{3}{l}{\textit{ADM Parameters}} \\
\hline
B-dipole bending angle & 0.26 & rad \\
B-dipole radius & 10 & cm \\
Dogleg bending angle & 0.2 & rad \\
Dogleg radius & 5.46 & cm \\
Drift length & 5 & cm \\
THz modulation depth & 0.12 & \% \\
THz wavelength & 1000 & $\mu$m \\
\bottomrule
\end{tabular}
\end{table}

\begin{figure}[htbp]
    \centering
    \subfloat[ ]{\includegraphics[width=0.49\linewidth]{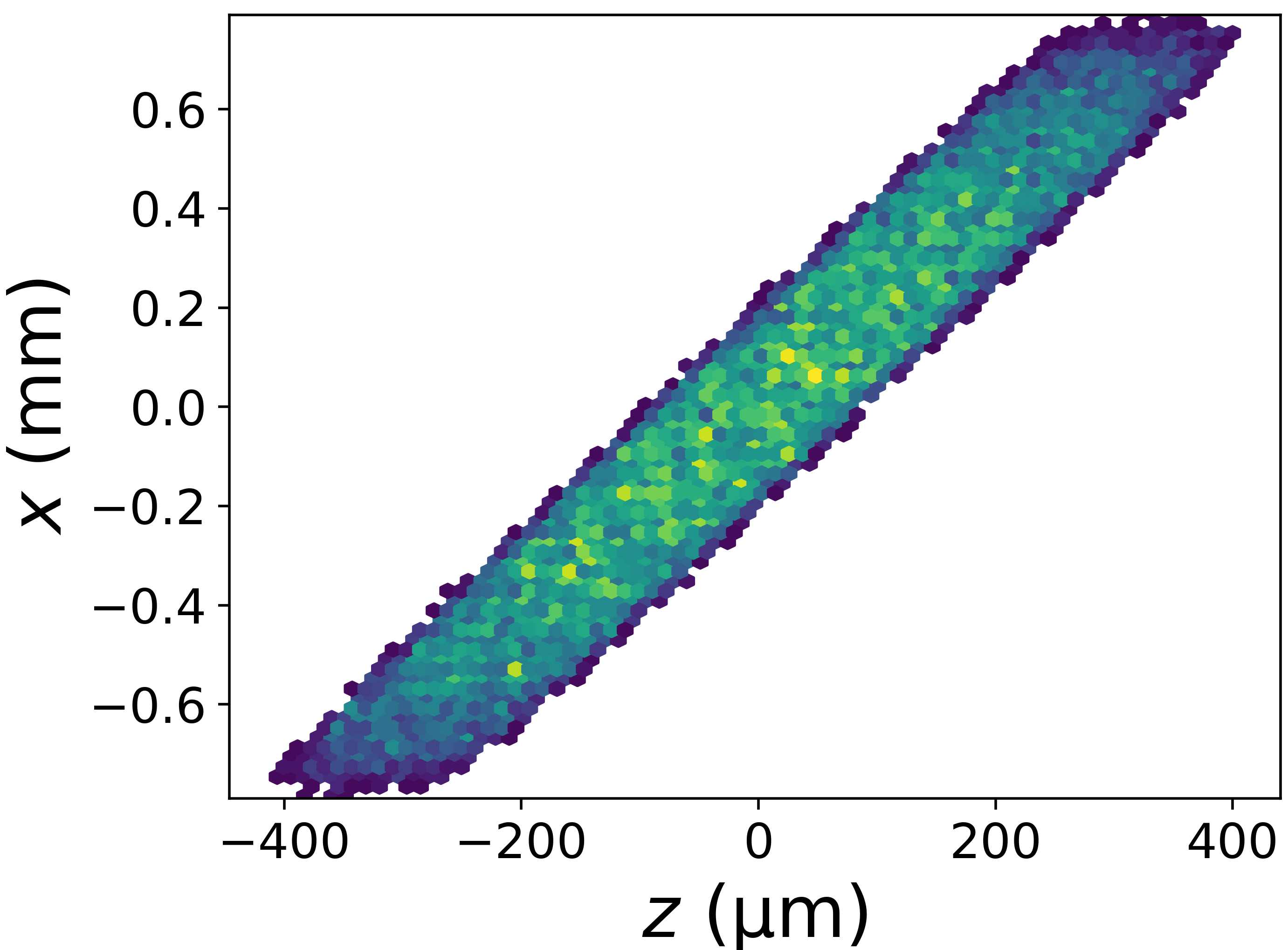}\label{fig:zx_preB}}
    \hfill
    \subfloat[]{\includegraphics[width=0.49\linewidth]{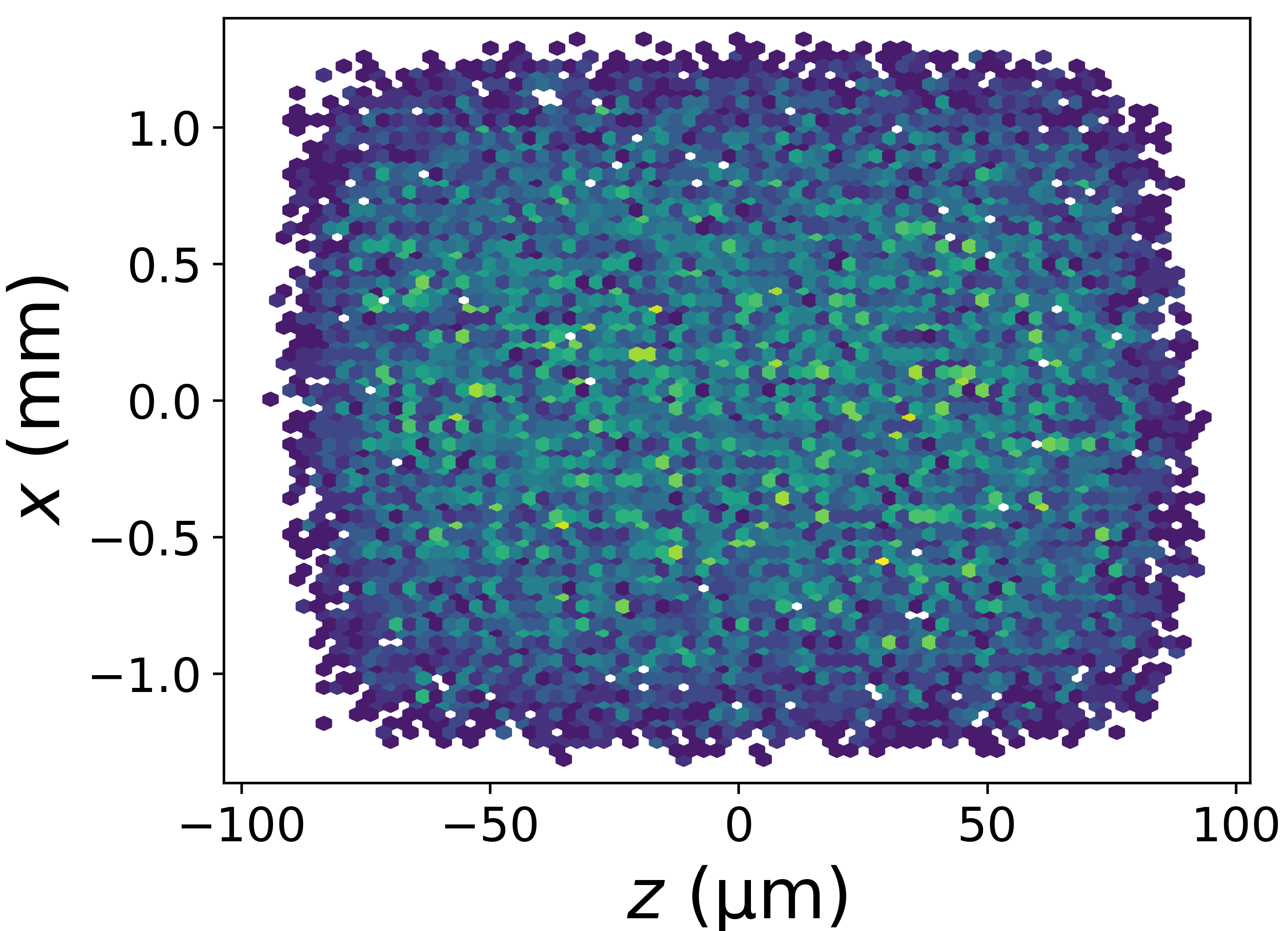}\label{fig:zx_postB}} \\
    
    \subfloat[]{\includegraphics[width=0.49\linewidth]{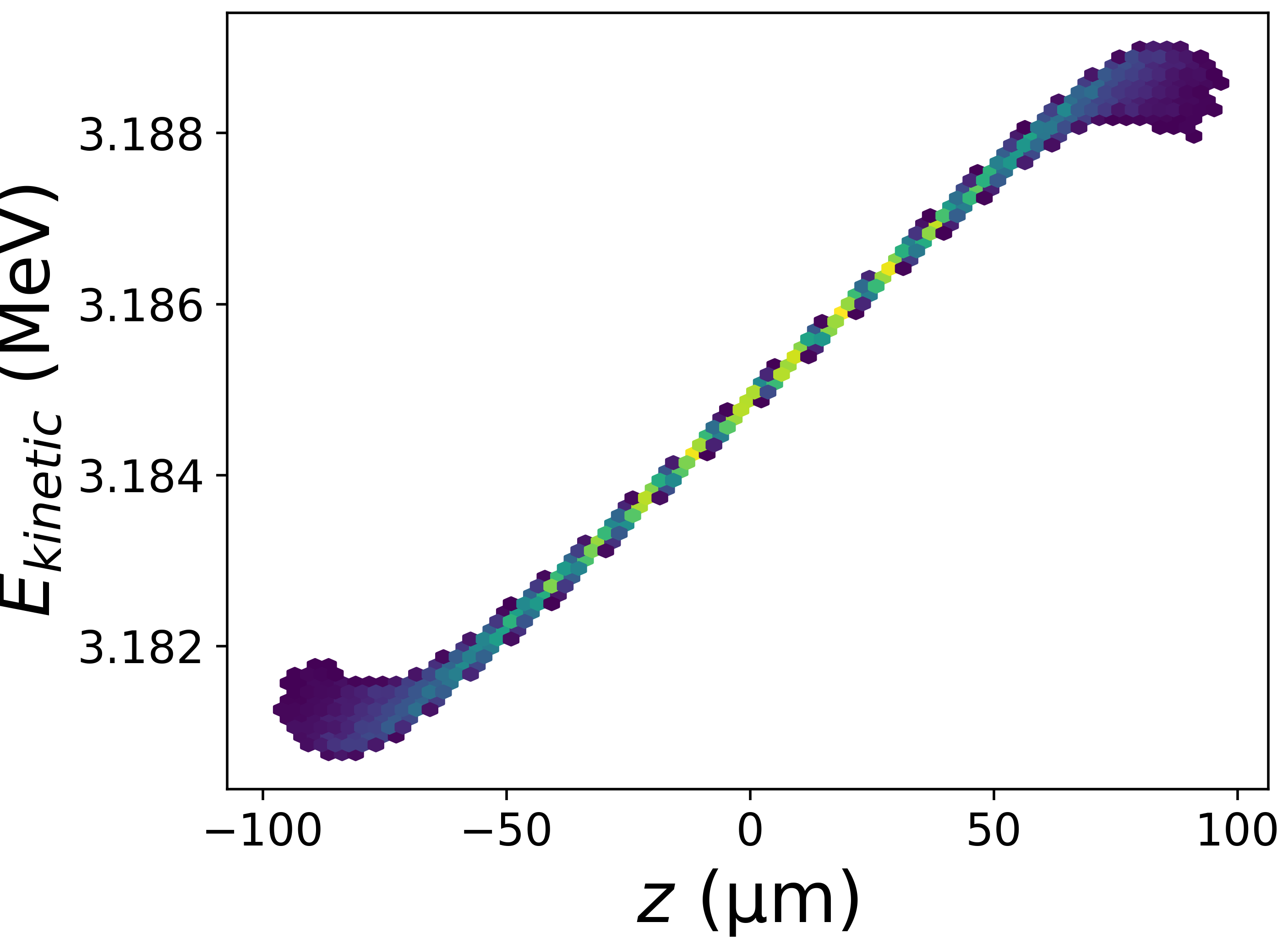}\label{fig:zE_postB_new}}
    \hfill
    \subfloat[]{\includegraphics[width=0.49\linewidth]{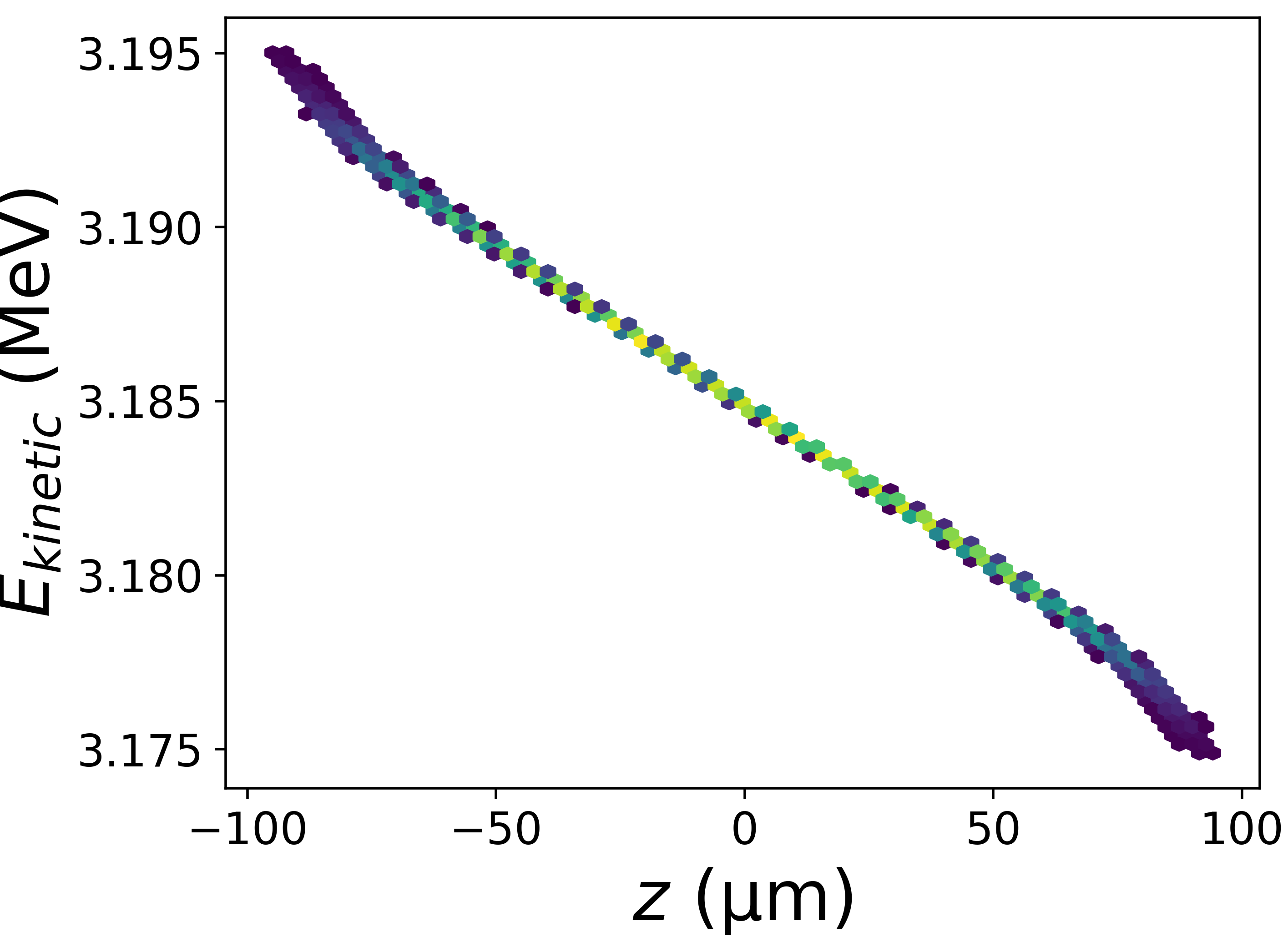}\label{fig:zE_postTHz}} \\
    
    \subfloat[]{\includegraphics[width=0.49\linewidth]{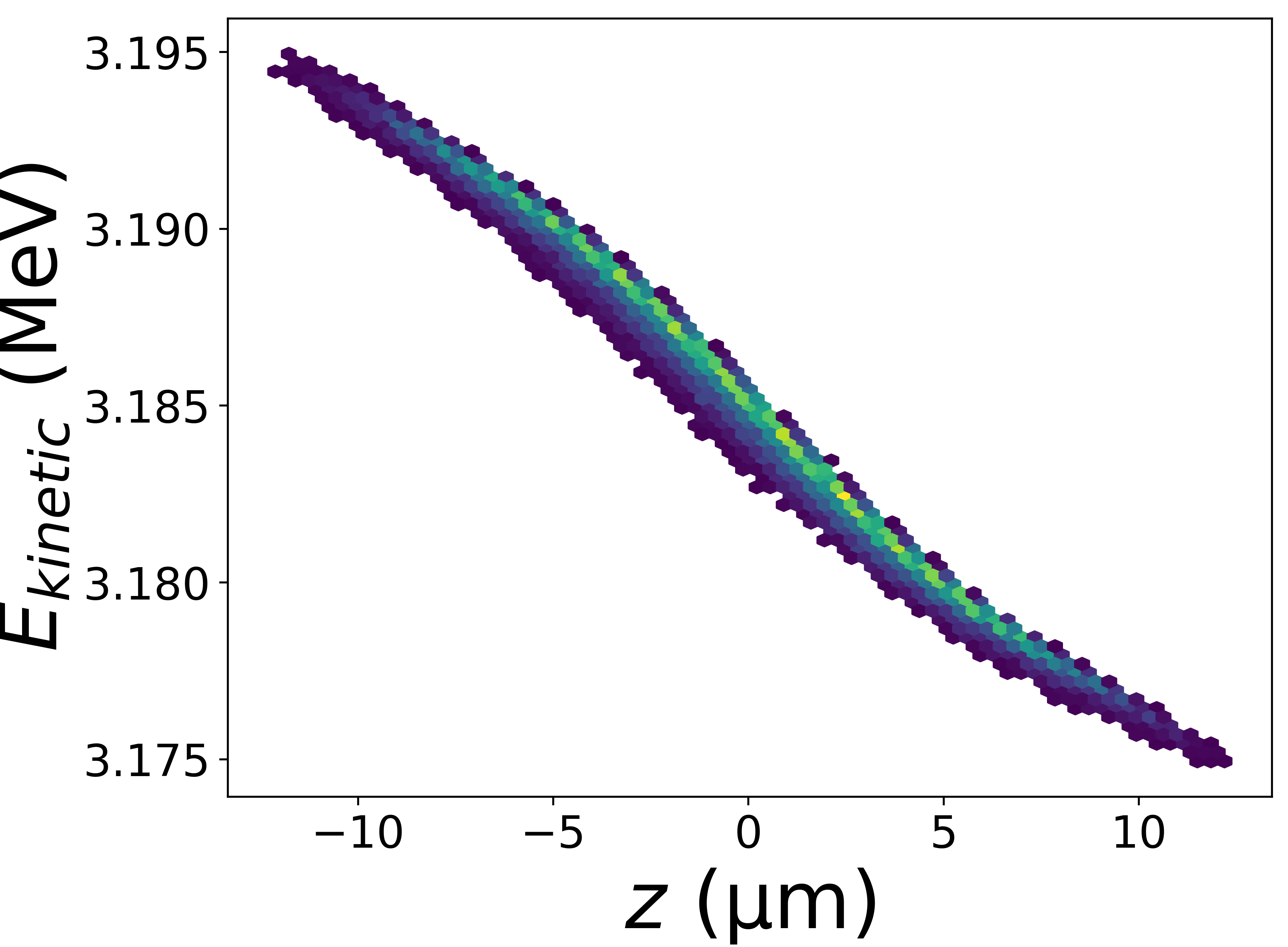}\label{fig:zE_postDogleg}}
    \hfill
    \subfloat[]{\includegraphics[width=0.49\linewidth]{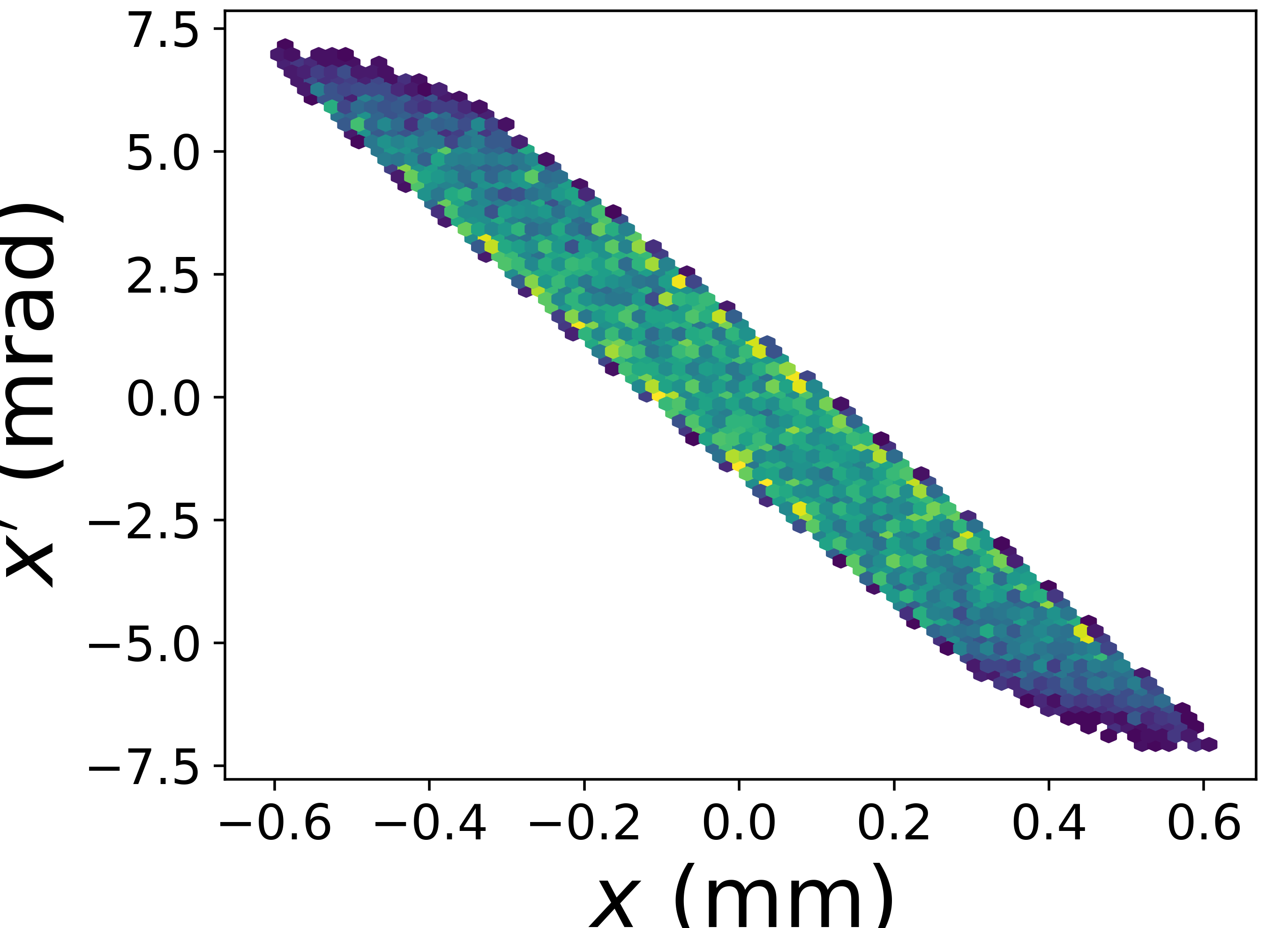}\label{fig:xxp_postDogleg}}
  \caption{Multi-dimensional phase-space manipulation in the beamline: 
    (a) Initial $z$-$x$ phase space correlation established via pulse-front-tilt laser illumination at the photocathode.
    (b) $z$-$x$ phase space post-dipole bending.
    (c) $z$-$\delta$ correlation post-dipole bending. 
    (d) THz-driven chirp polarity reversal ($\delta \propto -z$) through energy modulation.
    (e) Longitudinal density modulation after the dogleg. 
    (f) Transverse phase space post-dogleg.}
    \label{fig:phase space evolution}
\end{figure}
A UV laser pulse with a controlled pulse-front tilt irradiates the photocathode surface, establishing the pre-engineered correlation between the longitudinal position $z$ and transverse coordinate $x$ of emitted electrons. This pre-engineered $z$-$x$ coupling arises from the tilted laser wavefront interacting with the cathode geometry, effectively achieving transverse-longitudinal coupling for the electron bunch, as shown in Fig.~2a.

Following the electron beam is accelerated in a 2.4-cell S-band RF gun operating at 2856 MHz with a peak accelerating gradient of 63 MV/m. Unlike conventional 1.6-cell RF guns~\cite{qi_breaking_2020}, this extended cavity design suppresses higher-order longitudinal dispersion terms ($R_{56}^{\rm gun}$) that typically distort the linear energy chirp in compact cavities. By leveraging space-charge-driven beam expansion during initial propagation, the gun deterministically imprints a linear energy chirp while maintaining centroid energy stability. Crucially, the extended cavity geometry enables precise control and analytical abstraction of the longitudinal dispersion $R_{56}^{\rm gun}$ through this uniquely designed 2.4-cell configuration, which effectively suppresses higher-order effects and isolates a stable $R_{56}^{\rm gun}$ value. Traditional RF guns, constrained by truncated cavity structures, suffer from dominant higher-order dispersive terms and fail to produce a stable $R_{56}^{\rm gun}$. The ability to isolate and engineer $R_{56}^{\rm gun}$ provides a foundational degree of freedom for enforcing the isochronicity condition $R_{56}^{\rm total} = 0$ (Eq. 1), ensuring a robust temporal compression independent of RF amplitude noise.

The accelerated beam then enters the first bending dipole of ADM, a critical component that introduces transverse angular dispersion. In conventional systems, such a dipole would couple transverse divergence (\(x'\)) to the longitudinal displacement (\(z\)), stretching the bunch and generating nonlinear phase space distortions. This occurs because electrons with different energies follow distinct trajectories through the dipole, translating energy spread into transverse positional spread. However, as shown in Fig.~2a, the pre-established \(z\)-\(x\) correlation from oblique laser illumination provides a geometric counterbalance to this coupling. The dipole redistributes electrons horizontally according to their energy deviations (\(x' \propto \delta\)), but the initial \(z\)-\(x\) mapping ensures that longitudinal position remains linearly correlated with transverse coordinate, as depicted in Fig.~2b. This pre-compensation mechanism suppresses the dipole's natural tendency to elongate the bunch and preserves phase space linearity, enabling uniform energy modulation in subsequent stage.

\begin{figure}[htbp]
\centering
\begin{subfigure}{0.49\linewidth}
\includegraphics[width=\linewidth]{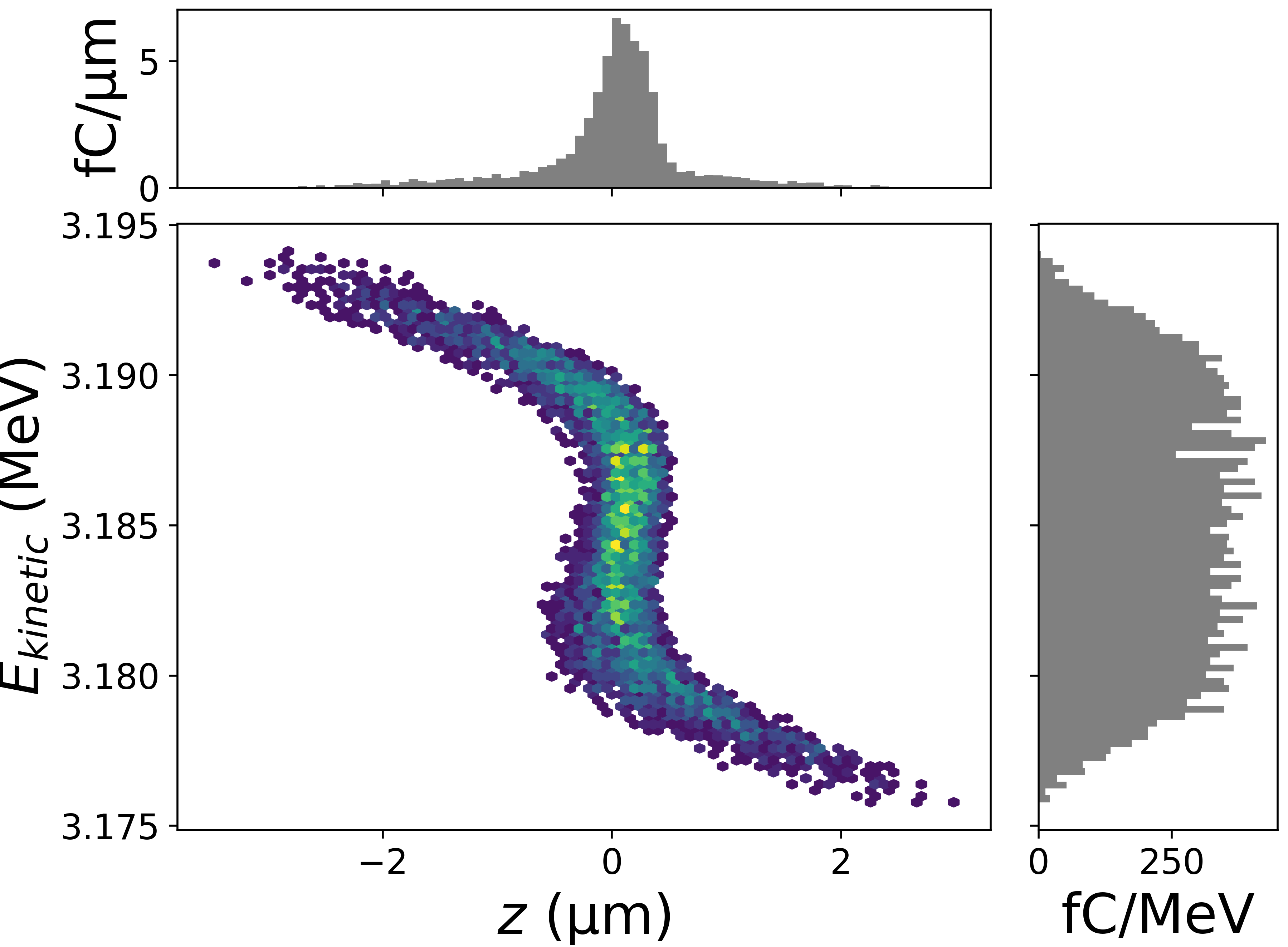}
\caption{}
\label{fig:adm_phase}
\end{subfigure}
\hfill
\begin{subfigure}{0.49\linewidth}
\includegraphics[width=\linewidth]{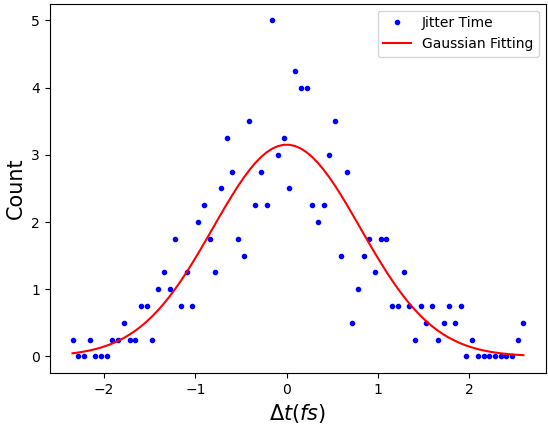}
\caption{}
\label{fig:adm_jitter}
\end{subfigure}
\caption{Final compression results with the multi-dimensional phase-space manipulation method: (a) Longitudinal phase space distribution and and (b) arrival time jitter distribution.}
\label{fig:adm_results}
\end{figure}

Following dipole-induced phase space conditioning, the electron beam enters the THz cavity and acquires energy modulation at the zero-crossing of the 0.3 THz field. As depicted in Fig.~2c, the electron beam initially carries a positive energy chirp (\(h = +15\ \mathrm{keV/ps}\)). The THz interaction then imparts a spatially uniform energy shift \(\delta = -h_{\mathrm{THz}} z\) with \(h_{\mathrm{THz}} = -43\ \mathrm{keV/ps}\), thereby flipping the chirp to \(h = -28\ \mathrm{keV/ps}\), as illustrated in Fig.~2d. Next, the dogleg compressor performs the transverse-to-longitudinal emittance partitioning required for temporal compression, as shown in Fig.~2e and f. After the final drift, the THz-imposed chirp is fully converted into bunch shortening, yielding an 810~as (rms) pulse at the sample plane while retaining 6~fC of charge post-collimation. This sequence realizes the phase-space transformation of Eq. (3), in which the transverse dispersion maps divergence \(x'\) into longitudinal displacement \(z_{\rm final} = \eta x'\), as shown in Fig.~\ref{fig:adm_phase}. The bunch duration has been compressed to 810~as (rms) at the sample plane, with the charge maintained at 6~fC after collimation. Within the FWHM window, 76.4\% of the electrons were retained, demonstrating an efficient signal-to-noise ratio. 

To demonstrate the stability and reproducibility of the proposed method, we simulated 100 consecutive electron bunches for arrival time jitter analysis. Following experimentally achievable parameter specifications, the simulations incorporated the following jitter sources: 0.05\% root-mean-square (rms) RF amplitude fluctuations, 0.2~ps rms phase jitter, 3\% charge fluctuations, and 0.01\% rms magnetic field strength variations. Figure~\ref{fig:adm_jitter} quantifies the shot-to-shot temporal stability, revealing 850~as rms arrival time jitter across 100 consecutive shots.

This multi-dimensional phase-space manipulation departs fundamentally from conventional \(R_{56}\)-based compression. Instead of relying solely on longitudinal dispersion (\(R_{56}\)) to convert energy spread into temporal compression, the ADM scheme exploits transverse angular dispersion (\(\eta\)) as its primary driver. By redistributing beam parameters across transverse and longitudinal dimensions, it deliberately trades transverse emittance for far tighter longitudinal compression. Under identical beam and THz-modulation conditions, simulations have been perfomed and the results show that a traditional \(R_{56}\) compressor achieves only 9.2~fs (rms) bunch duration. 

Finally, following the above temporal compression, the electron beam develops transverse emittance anisotropy due to the asymmetric phase-space partitioning in the ADM section. The horizontal and vertical phase-space distributions decouple, yielding a flat beam profile with increased horizontal divergence. To restore transverse symmetry, a skewed quadrupole doublets (Q1, Q2) are placed downstream of the dogleg to introduce controlled coupling between the horizontal and vertical planes. By redistributing horizontal divergence into the vertical plane, they equalize the rms beam sizes in both dimensions~\cite{Kim:2023wub}. Simulations confirm that this active emittance control transforms the flattened profile into a round beam with horizontal and vertical rms sizes of approximately 90 $\mu m$ at the sample.

\begin{figure}[htbp]
\centering
\begin{subfigure}{0.49\linewidth}
\includegraphics[width=\linewidth]{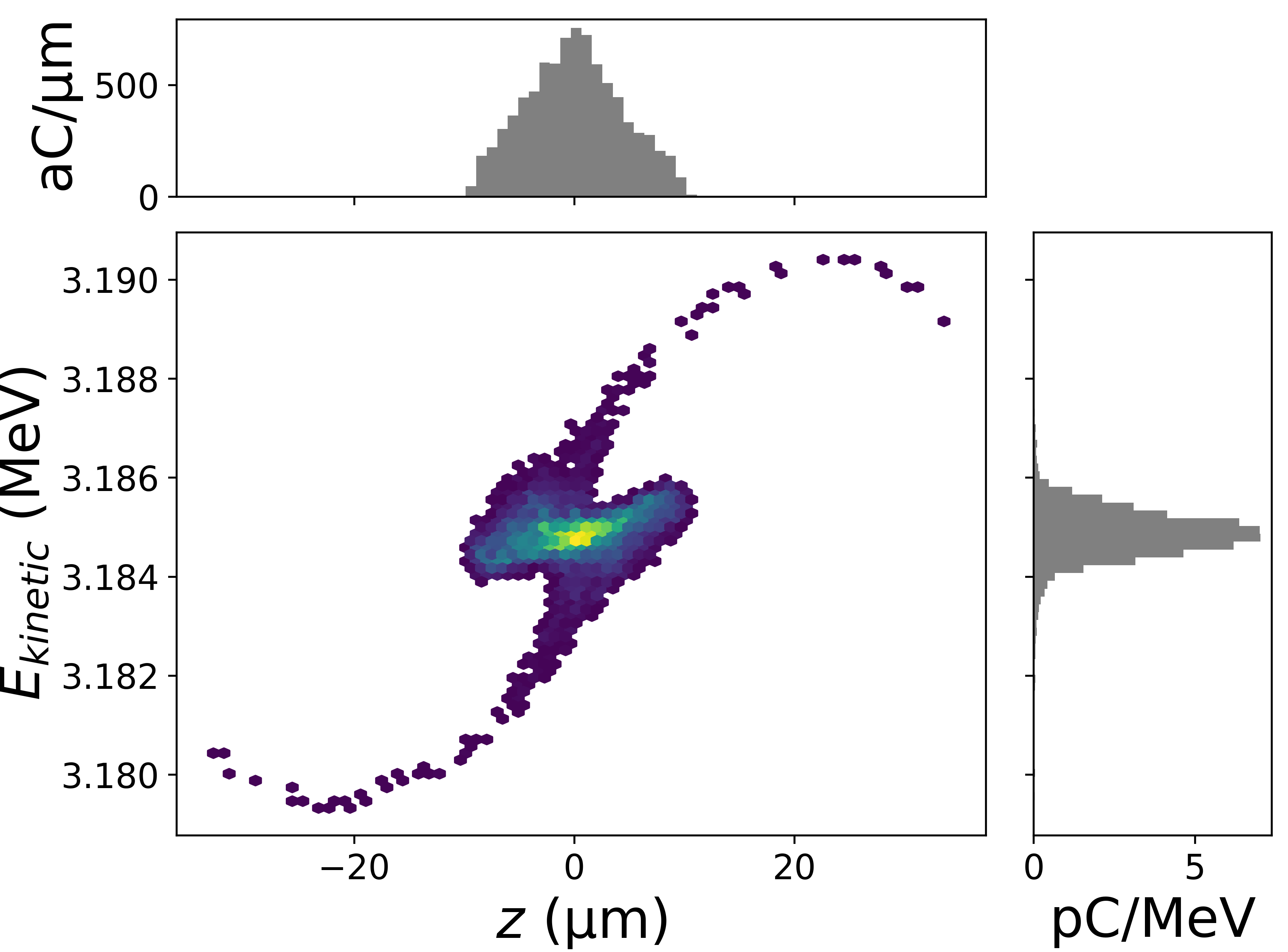}
\caption{}
\label{fig:dba_phase}
\end{subfigure}
\hfill
\begin{subfigure}{0.49\linewidth}
\includegraphics[width=\linewidth]{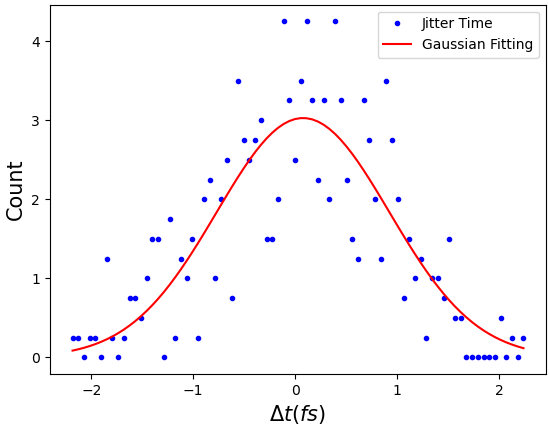}
\caption{}
\label{fig:dba_jitter}
\end{subfigure}
\caption{DBA compression results: (a) Longitudinal phase space distribution and and (b) arrival time jitter distribution.}
\label{fig:dba_results}
\end{figure}

To better illustrate the compression capability of the ADM scheme, we performed a comparative study with the DBA scheme under identical electron bunch parameters. DBA compression scheme achieves temporal focusing by counterbalancing space-charge-induced energy chirp with longitudinal dispersion~\cite{qi_breaking_2020}. Figure 4(a) illustrates the longitudinal phase space evolution through the DBA system. Numerical simulations using parameters from Table~\ref{tab:full_params} demonstrate that a 50 fs initial beam can be compressed to 14 fs (rms) (Fig.~\ref{fig:dba_phase}), with the charge maintained at 17~fC after compression, retaining 77.8\% of electrons within the central temporal window. The residual timing jitter is about 880~as (rms) across 100 consecutive shots (Fig.~~\ref{fig:dba_jitter}).

The DBA's performance degradation at higher charges stems from space-charge-driven phase space rotation. As shown in Fig.~4(a), space charge forces continuously accelerate head electrons and decelerate tail electrons, inducing clockwise rotation in longitudinal phase space. This rotation counteracts the designed $R_{56}$-based compression, reducing effective chirp linearity.

\section*{Summary}  
In summary, a novel multi-dimensional phase-space manipulation approach has been proposed to realize robust attosecond-scale compression of MeV electron beams while suppressing timing jitter. By combining pulse-front-tilted laser excitation at the photocathode for transverse-to-longitudinal mapping, THz-driven linear energy chirping immune to RF jitter, and a transverse-longitudinal emittance partitioning beamline that converts transverse emittance into longitudinal compression, a 3 MeV electron beam is compressed from 50 fs to 810 as (rms) with 850 as arrival-time jitter, retaining 6 fC of the charge. This represents an order-of-magnitude improvement in both pulse duration and timing stability over state-of-the-art two-dimensional manipulation schemes. This multi-dimensional paradigm opens the door to attosecond-resolved investigations of atomic and electronic dynamics in ultrafast science, and has broad applicability to other accelerator-based sources requiring ultrashort electron beams.  

\begin{acknowledgments}
The authors would like to thank Y. Kang, Y. Liu and J. Xiong for helpful discussions. This work was supported by the National Natural Science Foundation of China (Grant No.12435011), Project for Young Scientists in Basic Research of Chinese Academy of Sciences (YSBR-115), and Shanghai Municipal Science and Technology Major Project.
\end{acknowledgments}

\bibliographystyle{ieeetr}
\bibliography{main}

\end{document}